\documentclass[10.75pt, a4paper]{article}

\usepackage[utf8]{inputenc}
\usepackage[]{parskip} 
\usepackage[tbtags]{mathtools}
\usepackage{amsmath,amssymb,siunitx,bm,nicefrac}
\let\vec\mathbf
\usepackage{graphicx}
\usepackage[font=footnotesize]{caption}
\usepackage[dvipsnames,x11names]{xcolor}
\usepackage{orcidlink}
\usepackage{url}

\usepackage{geometry}
\usepackage{titlesec}
\titleformat{\section}
  {\bf\sffamily}
  {\thesection. }
  {5pt}
  {\MakeUppercase}
\renewcommand{\thesection}{\Roman{section}} 

\titleformat{\subsection}
  {\bf\sffamily}
  {\thesubsection. }
  {5pt}{}
  
\renewcommand{\thesubsection}{\Alph{subsection}}
\usepackage[affil-it]{authblk} 
\usepackage{etoolbox}
\usepackage{lmodern}

\usepackage{hyperref}
\hypersetup{
   hidelinks,
   colorlinks,
   linkcolor={blue!50!black},
   citecolor={blue!50!black},
   urlcolor={blue!80!black}
}

\usepackage[citestyle=numeric,style=phys,biblabel=brackets,backend=bibtex,sorting=none,url=false,doi=false, natbib]{biblatex}

\DefineBibliographyStrings{english}{andothers={\itshape et\addabbrvspace al\adddot}}

\makeatletter
\patchcmd{\@maketitle}{\LARGE \@title}{\fontsize{16}{19.2}\selectfont\@title}{}{}
\makeatother

\usepackage{abstract}

\title{Chirality inversion and odd transport in a suspension of living rotors} 

\author[1]{Isabelle Eisenmann\thanks{i.eisenmann@uva.nl, ORCID: \orcidlink{0009-0000-2778-0930}}}
\author[1]{Pedro Cosme}
\author[1]{Hannah van Sterkenburg}
\author[1]{Hugo Fran\c{c}a}
\author[1]{Titouan Seguy}
\author[1]{Maria T\u{a}tulea-Codrean}
\author[1,2]{Maziyar Jalaal\thanks{m.jalaal@uva.nl;m.jalaal@damtp.cam.ac.uk, ORCID: \orcidlink{0000-0002-5654-8505}}}

\affil[1]{Van der Waals-Zeeman Institute, Institute of Physics,
University of Amsterdam,
\newline Science Park 904, 1098 XH Amsterdam, The Netherlands}

\affil[2]{Department of Applied Mathematics and Theoretical Physics,
University of Cambridge,
\newline Wilberforce Road, Cambridge CB30WA, United Kingdom}

\begin{document}

\begingroup
\sffamily
\date{}
\maketitle
\endgroup

\begin{abstract}

Chiral symmetry breaking is a defining feature of life, visible at all scales from molecules to organisms. Yet how microscopic chirality propagates to shape the macroscopic properties of living matter remains poorly understood.
Here, using a new experimental system of spinning microalgae, we show that a living chiral fluid behaves fundamentally differently from an externally driven one.
As their density increases, cells self-assemble hydrodynamically into small rotating clusters which merge into a percolating chiral fluid. Strikingly, microscopic and macroscopic vorticity have opposite signs: individual cells all spin clockwise, yet clusters orbit counterclockwise. Combining experiments, theory and simulations, we show that both chirality inversion and percolation are direct consequences of the cells' internally driven nature. We then show how percolation reshapes transport, with even and odd diffusivities responding differently to the emerging network topology.


\end{abstract}

Active matter has traditionally distinguished active from passive systems through self-propulsion, associated with the breaking of time-reversal symmetry by each particle \cite{marchetti_hydrodynamics_2013, obyrne_time_2022}. A defining feature of living systems, however, is the additional breaking of parity, or mirror symmetry \cite{pasteurDissymetrieMoleculaireConference, badaMarineSedimentsDating1970}. Life is built from homochiral molecules: DNA and RNA almost exclusively form right-handed helices, while proteins are composed of left-handed amino-acids. This molecular asymmetry propagates to all larger scales, up to cells, tissues and organisms - as visible for instance in the left-sided location of our heart
\cite{lebretonMolecularOrganismalChirality2018, mcgrathTwoPopulationsNode2003}.
Materials that break both time-reversal and parity symmetry at the microscopic level exhibit new macroscopic behaviors such as odd elasticity, viscosity and diffusivity, where the response is orthogonal to the driving force \cite{avronOddViscosity1998, fruchartOddViscosityOdd2023, hargusFluxHypothesisOdd2025}. 
Ensembles of particles spinning in the same direction provide the simplest examples of such chiral active matter \cite{liebchenChiralActiveMatter2022}.
Experimental realizations remain scarce: colloidal magnets in a rotating magnetic field spin because an external torque acts on each particle: they are driven \cite{grzybowskiDynamicSelfassemblyMagnetized2000, soniOddFreeSurface2019, meckeSimultaneousEmergenceActive2023}. Biological rotors are self-actuated, but those studied so far crystallize instead of forming a fluid \cite{petroffFastMovingBacteriaSelfOrganize2015, drescherDancingVolvoxHydrodynamic2009, tanOddDynamicsLiving2022};
a fluid of particles that generate their own rotation is still missing.

Here, we bridge the gap by introducing a novel experimental system of living chiral fluid, and show that it differs fundamentally from its externally driven counterparts. We use a uniflagellar mutant of the model microalga \textit{Chlamydomonas reinhardtii} as a living rotor: with a single flagellum, the cells cannot swim but instead spin clockwise in place, breaking both parity and time-reversal symmetry. They interact hydrodynamically and self-assemble into highly dynamic rotating clusters, which percolate as the cell packing fraction increases. Strikingly, we observe an inversion of chirality between microscopic and macroscopic scales: individual cells spin clockwise, yet clusters orbit counterclockwise. The same reversal appears in transport, with the odd diffusivity of sign opposite to the spinning direction of the cells. Combining experiments, theory and simulations, we show that these collective behaviors stem from the internally driven nature of the cells: because they exert no net torque on the fluid, their rotation is balanced by a counter-rotating spiral flow that advects neighboring cells in the opposite direction while drawing them together. This contrasts with driven rotors, which generate a simple vortex, and highlights how active and driven systems can give rise to different collective phenomenologies.
In the following, we first examine the dynamics of a single spinning cell before building up to collective properties, with a focus on odd transport.

\subsection*{A living rotor with fixed chirality: the \textit{C. reinhardtii} uni1 mutant}

\begin{figure}
	\includegraphics[width=1\columnwidth]{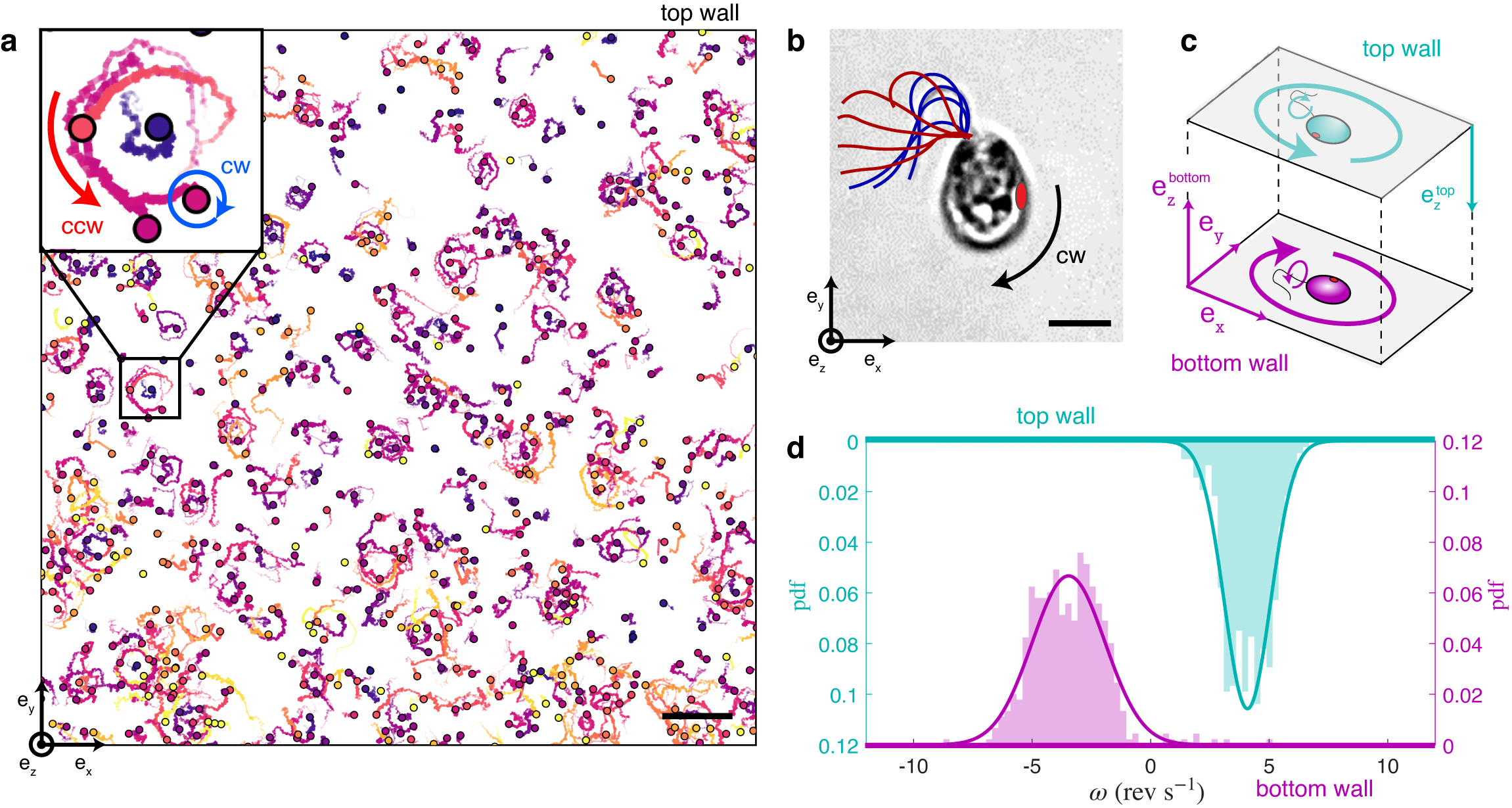} 
	\caption{\textbf{The uniflagellar mutant of \textit{C. reinhardtii}}.
        \textbf{a}: Collective dynamics of a cell suspension, with clusters of cells rotating with a macroscopic vorticity (CCW) opposite to single cells' (CW). Scale bar \SI{100}{\micro\meter}. Color-coded by speed, blue to yellow.
		\textbf{b}: Single cell with eyespot (red spot) and the \SI{50}{\hertz} flagellum beating cycle superimposed (power stroke in red, recovery stroke in blue). Cells all spin CW near a wall. Scale bar \SI{10}{\micro\meter}.
		\textbf{c}: Spinning directions of sedimenting and floating cells are opposite when viewed with the microscope, but are consistent with regard to the normal pointing from the wall to the fluid.
		\textbf{d}: Histogram of cells' spinning speed at the top and bottom walls of the chamber.
		\label{fig:Fig1}
	}
\end{figure}

To understand how clockwise-spinning cells self-assemble into a percolating fluid with macroscopic counterclockwise chirality, we start at the single-cell level.
Our microrotors are uniflagellar mutants of the model microalga \textit{C. reinhardtii} \cite{huangUniflagellarMutantsChlamydomonas1982, baylyPropulsiveForcesFlagellum2011}. Developed to study the biology of the flagellum, they have so far barely been used in the context of active matter. Crucially, these mutants all retain the same flagellum --- the \textit{trans}, located opposite to their eyespot, and therefore have the same intrinsic chirality (Fig.~\ref{fig:Fig1}b). Cells are loaded in a microfluidic chamber of \SI{100}{\micro\meter} height, much larger than their radius $a\simeq\SI{5}{\micro\meter}$. With a single flagellum, they are unable to swim and most sink to the bottom of the chamber, where they rotate in place. Remarkably, essentially all cells rotate in the same direction: clockwise when viewed from the fluid, at a frequency of about \SI{4}{Hz}. This shared handedness is the defining feature making a suspension of these cells a chiral fluid, breaking both parity and time-reversal symmetry (Fig.~\ref{fig:Fig1}a).
It originates from the preferential orientation of the cells against a wall: if they were equally likely to settle on either their ``ventral'' or ``dorsal'' side, both clockwise (CW) and counterclockwise (CCW) rotation would be observed. Effectively, a small out-of-plane component of the flagellum beat (responsible for the left-handed helical trajectories observed in wild-type cells \cite{corteseControlHelicalNavigation2021}) induces a weak left-handed roll of the cells around their main axis. As a result, cells lay against the wall with their flagellum on the left when viewed from the fluid, where they rotate almost exclusively CW (Fig.~\ref{fig:Fig1}c).

A subpopulation of lower-density cells can be found floating below the upper wall of the chamber. There, the same mechanism acts in a mirrored wall geometry, causing them to rotate CCW when viewed from above (Fig.~\ref{fig:Fig1}d) --- an effect also observed in bacteria \cite{diluzioEscherichiaColiSwim2005, beppuEdgeCurrentPairing2021}. In the following, we describe experiments performed either at the top or bottom wall of the chamber, and always restricted to a single plane. Cells at the top and bottom walls are too far apart to interact, as explained below. For simplicity, in both cases we define the direction normal to the wall pointing toward the fluid as $+\mathbf e_z$. With this convention, cells always rotate CW at a frequency $\simeq$\SI{4}{\hertz} (Fig.~\ref{fig:Fig1}c).

\subsection*{Hydrodynamic model of a single chiral cell}

\begin{figure}[ht!]
	\includegraphics[width=1\columnwidth]{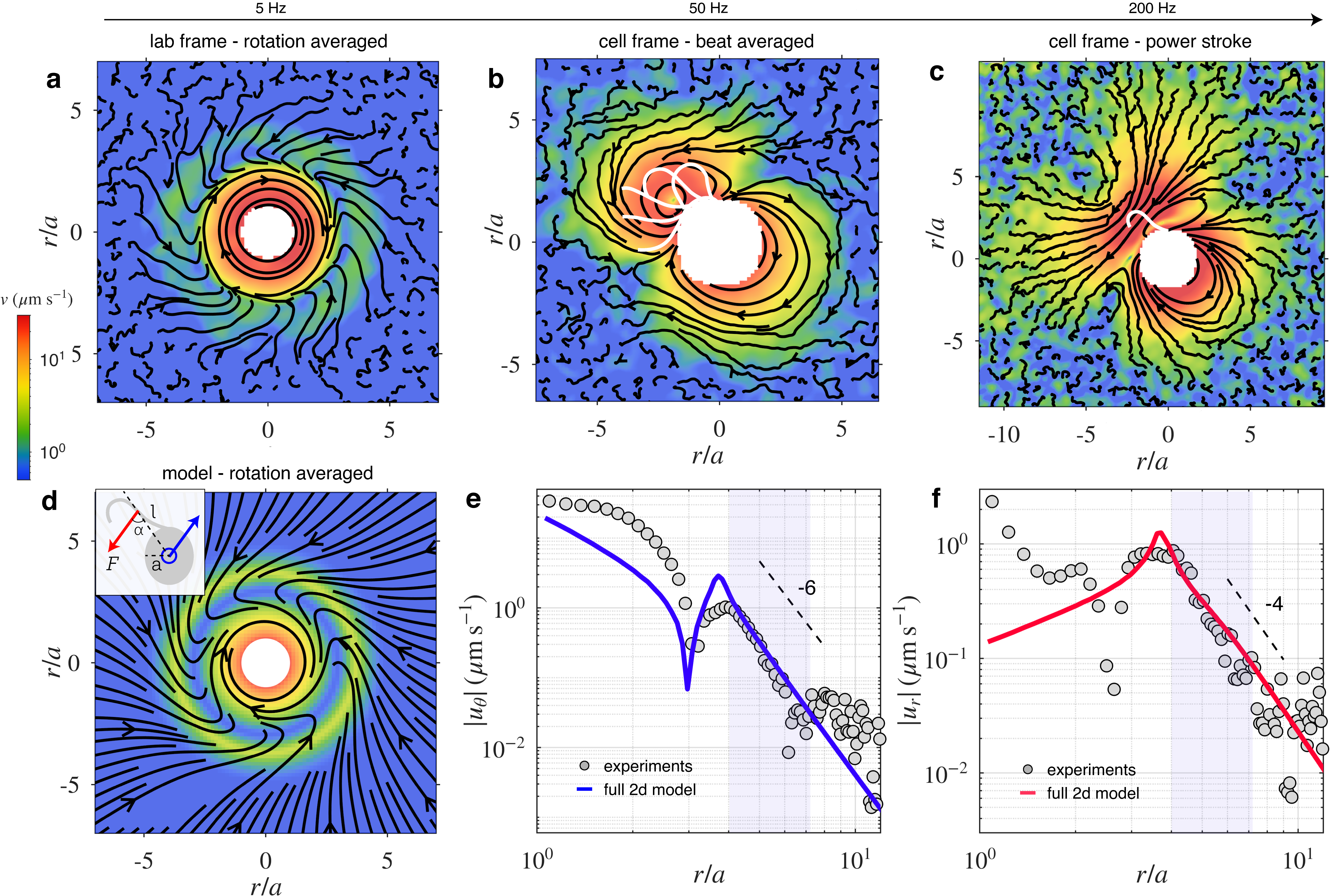} 
	\caption{
        \textbf{Hydrodynamic model of a single cell}.
        Experimental flow field averaged over cell rotation in the laboratory frame (\textbf{a}), averaged over one flagellar beat cycle in the cell frame (\textbf{b}) and measured during the power stroke of the beat (30\% highest body angular speeds, \textbf{c}). In panels b and c, only round cells of ellipticity $>0.95$ are considered, and cell orientation is rectified prior to averaging.
        \textbf{d}: Flow field resulting from the superposition of two Stokeslets and one rotlet, averaged over rotation around the origin (with $F_{cycle}=\SI{1}{\pico\newton}$, $\alpha_{cycle}=\SI{126}{\degree}$, $\ell_{cycle}=\SI{18}{\micro\meter}$, $a=h=\SI{5}{\micro\meter}$). Inset: model of a single cell. Arrows represent the forces and the circle represents the torque.
        \textbf{e, f}: Radial decay of the azimuthal and radial components of the experimental rotation-averaged flow field, compared with the model in panel d with manually fitted parameters. The shaded area represents the equilibrium cell-cell distance. $u_\theta$ changes sign at $r\simeq3a$.               
	}
    \label{fig:Fig2}
\end{figure}

As interactions are mediated by the fluid, we first examine the flow field around individual cells by particle tracking velocimetry of passive tracers introduced into the suspension. 
The single-cell dynamics is set by two well-separated timescales, with the flagellar beat (\SI{50}{\hertz}) powering the slower rotation of the cell body (\SI{4}{\hertz}). To resolve the beat itself, we image the tracers at \SI{200}{\hertz}. Averaging tracer trajectories over appropriate time windows then yields complementary views of the flow field (Fig.~\ref{fig:Fig2}a-c).
In the laboratory frame and averaged over its rotation, a cell generates a striking spiral flow: the fluid is drawn CCW toward the cell down to a distance $r=3a$ from its body center, where it switches direction (Fig.~\ref{fig:Fig2}a).  The fluid velocity decays relatively rapidly far from the cell and is not distinguishable from noise for $r>7a$.
This vorticity reversal is a direct consequence of the internally driven nature of the rotors. Indeed, at low Reynolds number and in the absence of external forcing, cells cannot exert a net force or torque on the surrounding fluid \cite{laugaHydrodynamicsSwimmingMicroorganisms2009}: in order to spin in one direction, they must push the fluid the other way. The counter-rotating vortices stem respectively from CW advection by the rotating body and the CCW cycle of the flagellum, as can be seen by averaging tracer data over one beat in the cell frame, which isolates both contributions (Fig.~\ref{fig:Fig2}b). Finally, the instantaneous flow field during the power phase of the beat reveals fluid pulling by the flagellum and additional side vortices (Fig.~\ref{fig:Fig2}c - recovery phase shown in SI).

Both the instantaneous and rotation-averaged velocity fields can be captured quantitatively by a minimal hydrodynamic model. Based on the work of Fily \textit{et al.} \cite{filyCooperativeSelfpropulsionActive2012}, we model the cell's contribution to the flow as the simplest realization of an active 2D rotor. Over a beating cycle, the flagellum exerts an effective active force $\mathbf{F}$ on the fluid, balanced by a drag $-\mathbf{F}$ acting on the cell body at a distance $l$, corresponding to the effective flagellum length. We denote by $\alpha$ the angle between $\mathbf{F}$ and the axis connecting the cell body centre to the average position of the flagellum (Fig.~\ref{fig:Fig2}d). The moment of these forces needs to be balanced by an active torque $\boldsymbol{\tau}$. 
Finally, since the cells are located near a wall, we account for boundary effects by using the method of images applied to each singularity \cite{blakeFundamentalSingularitiesViscous1974}. The instantaneous flow field follows from linear superposition of these contributions and can then be directly compared with experiments. Because isolated cells rotate essentially in place, following circular trajectories of radii smaller than their own size, we treat them as point spinners. Averaging over cell rotation, we obtain the following velocity field in the far-field limit:

\begin{equation}
	\mathbf{u}(r,\theta) = \frac{F}{8\pi \mu l}
	\left[
	\frac{12h^2l^2}{r^4}\cos\alpha\,\hat{\boldsymbol{r}}
	+
	\frac{45h^2l^4}{4r^6}\sin\alpha\,\hat{\boldsymbol{\theta}}
	\right]
    \label{eq:farfield}
\end{equation}

where $h$ is the vertical distance from the cell to the wall and $\mu$ the fluid viscosity ($\mu=\SI{1}{\milli\pascal\second}$).
This yields radial and azimuthal components decaying as $r^{-4}$ and $r^{-6}$ respectively, where $r$ is the in-plane distance from the cell. This steep decay results from both hydrodynamic screening by the wall and evaluating the three-dimensional flow in a two-dimensional plane, a geometric effect that is often overlooked. Indeed, while the wall changes the bulk velocities from $u_\rho\sim \rho^{-2}$ and $u_{\theta}\sim \rho^{-4}$ to $u_\rho\sim \rho^{-3}$ and $u_{\theta}\sim \rho^{-5}$, with $\rho$ the 3D distance to the cell \cite{filyCooperativeSelfpropulsionActive2012}, restricting the analysis to an observation plane at a finite distance from the no-slip boundary further steepens the decay of $u_{r}$ and $u_{\theta}$ by one order (see SI for details of the model).

This far-field model accurately captures the experimental flow field down to $r\sim3a$ from the cell center, both in the laboratory and cell frame (Fig.~\ref{fig:Fig2}d-f). Closer to the cell, the finite body size must be taken into account. Fitting experiments and model, we obtain $F=\SI{1}{\pico\newton}$ for the effective beat average thrust, and $F=\SI{12}{\pico\newton}$ for the power phase, consistent with previous measurements \cite{leptosPhototaxisChlamydomonasArises2023}. The recovery phase of the beat is also well described by this simple model (SI). Importantly, because steric interactions between their flagella prevent cells from approaching more closely, this far-field description is sufficient to understand hydrodynamic interactions, even in dense suspensions. 

To conclude this section, it is instructive to compare this system with other active and driven rotors. Here, because the flagellum beat is planar, both the vortex and counter-vortex appear in the observation plane, which underlies the chirality inversion observed at the collective level. This contrasts with spinning bacteria, where the body and flagellar torques are separated along the rotation axis, yielding a single apparent vortex around each cell body \cite{pedleyHydrodynamicPhenomenaSuspensions1992}. In larger spherical active rotors such as \textit{Volvox} colonies and starfish embryos (both actuated by thousands of cilia), torque balance is instead achieved through a compensation between azimuthal flow and body rotation, resulting in a flow field which appears purely radial in the laboratory frame \cite{tanOddDynamicsLiving2022, drescherDancingVolvoxHydrodynamic2009, pakGeneralizedSquirmingMotion2014}. In these systems, attraction is not intrinsic but comes from 3D recirculating flows due to the proximity of an interface. Finally, externally driven rotors (e.g. rotating colloidal magnets) generate purely azimuthal flows, with a single rotlet associated with the applied external torque, and therefore do not exhibit chirality inversion or attraction \cite{soniOddFreeSurface2019, meckeEmergentPhenomenaChiral2024}. We thus expect internally and externally driven microrotors to display qualitatively different collective behaviors, as will be shown below.

\subsection*{Spinning cells self-assemble into fluid-like clusters} 

\begin{figure}
	\includegraphics[width=1\columnwidth]{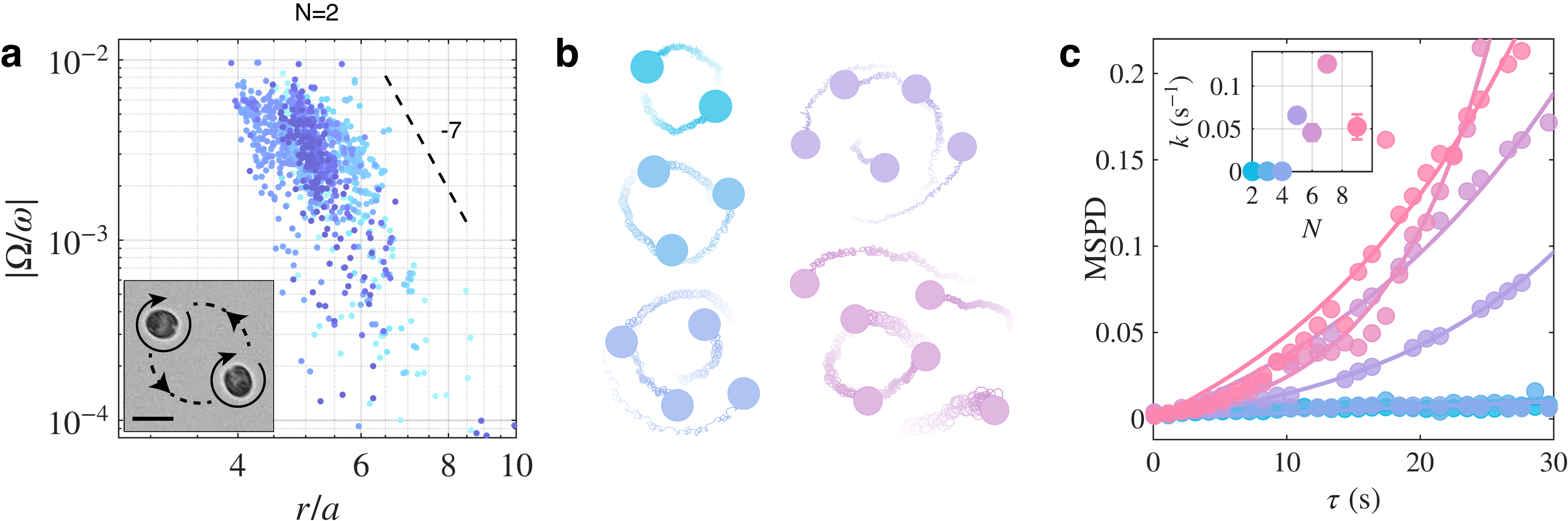} 
	\caption{\textbf{Melting of cell clusters}.
		\textbf{a}: Orbiting frequency $\Omega$ of cell pairs normalized by the mean individual spinning frequency $\omega$, versus cell separation $r/a$. Colors represent different pairs. Inset: pairs orbit CCW while cells spin CW (scale bar \SI{10}{\micro\meter}).
		\textbf{b}: Example trajectories of interacting cell groups of size $N=2-6$.
		\textbf{c}: Mean Squared Pairwise Displacement (MSPD) as a function of time lag $\tau$ for clusters of increasing size $N$ (8 replicates for $N=2$, 7 for $N=3$, 2 for $N=4$, 3 for $N=5$, and one otherwise. For $N$ with more than one replicate, each data point represents the average of at least two replicates at each time lag). Solid lines are fits of $k\tau$ for $N\leq4$ and $A(e^{k\tau}-1)$ for $N\geq5$. Inset: growth rate $k$ as a function of group size $N$. Error bars denote the 95\% confidence intervals of the fitted growth rates.
    }
    \label{fig:Fig3}
\end{figure}

Despite the rapid decay of the flow field, we observed cells to interact hydrodynamically. With the inward flow bringing them together, small rotating clusters of neighboring cells are frequently found, even in dilute suspensions. We now ask how collective structures build up, starting with such small groups. The simplest bound state consists of two cells orbiting each other. Strikingly, although individual cells spin CW, pairs orbit CCW (Fig.~\ref{fig:Fig3}a). Because the equilibrium separation between two cells $r_{eq}=4a-6a\simeq\SI{20}{\micro\meter}$ lies within the outer counter-rotating vortex they generate, neighboring cells are advected opposite to the direction of their individual spin. This chirality reversal contrasts with previous systems of interacting rotors and vortices \cite{grzybowskiDynamicSelfassemblyMagnetized2000, arefPointVortexDynamics2007,soniOddFreeSurface2019,tanOddDynamicsLiving2022, meckeSimultaneousEmergenceActive2023}, where the spinning and orbiting directions are the same.
Importantly, there is a pronounced separation of timescales between the cells individual spinning ($\sim\SI{4}{Hz}$), and collective orbiting frequencies ($\sim\SI{0.01}{Hz}$), which allows us to describe interactions using the rotation-averaged flow field. The measured orbital frequencies of cell pairs are consistent with the predicted azimuthal flow from (\ref{eq:farfield}), $\Omega = \frac{2u_\theta(r_{eq})}{r_{eq}} \sim r_{eq}^{-7}$ over the limited accessible range of distances between interacting cells
(Fig.~\ref{fig:Fig3}a).

Like pairs, rotating triplets and quadruplets appear to be robust to positional fluctuations and occasional obstacles. The high persistence of cell quadruplets in particular suggests that hydrodynamic attraction can stabilize configurations that would otherwise be chaotic \cite{arefIntegrableChaoticMotions1982, lushiPeriodicChaoticOrbits2015}. 
Larger clusters undergo frequent neighbor exchanges, marking the onset of a fluid-like state (Fig.~\ref{fig:Fig3}b). Given the small number of particles, we quantify this melting using the mean-squared pairwise displacement, defined as $\mathrm{MSPD}(\tau) = \frac{1}{2a^{2}} \left\langle \left[r_{ij}(t+\tau)-r_{ij}(t) \right]^{2} \right\rangle$, where $i,j$ are neighbors at time $t$ \cite{boThreePhasesOdd2026}. The MSPD shows negligible growth over the observation timescale for $N\leq4$, indicating structures rotating as a whole, but diverges for larger clusters, reflecting internal rearrangements (Fig.~\ref{fig:Fig3}c). Numerical simulations show that noise or cell polydispersity are essential for explaining this fluidization in the presence of attractive interactions, as detailed in SI. 
We therefore expect a dense suspension to behave as an active chiral fluid rather than an active solid.

\subsection*{Dense suspensions behave as a percolating chiral fluid}

\begin{figure}
	\includegraphics[width=1\columnwidth]{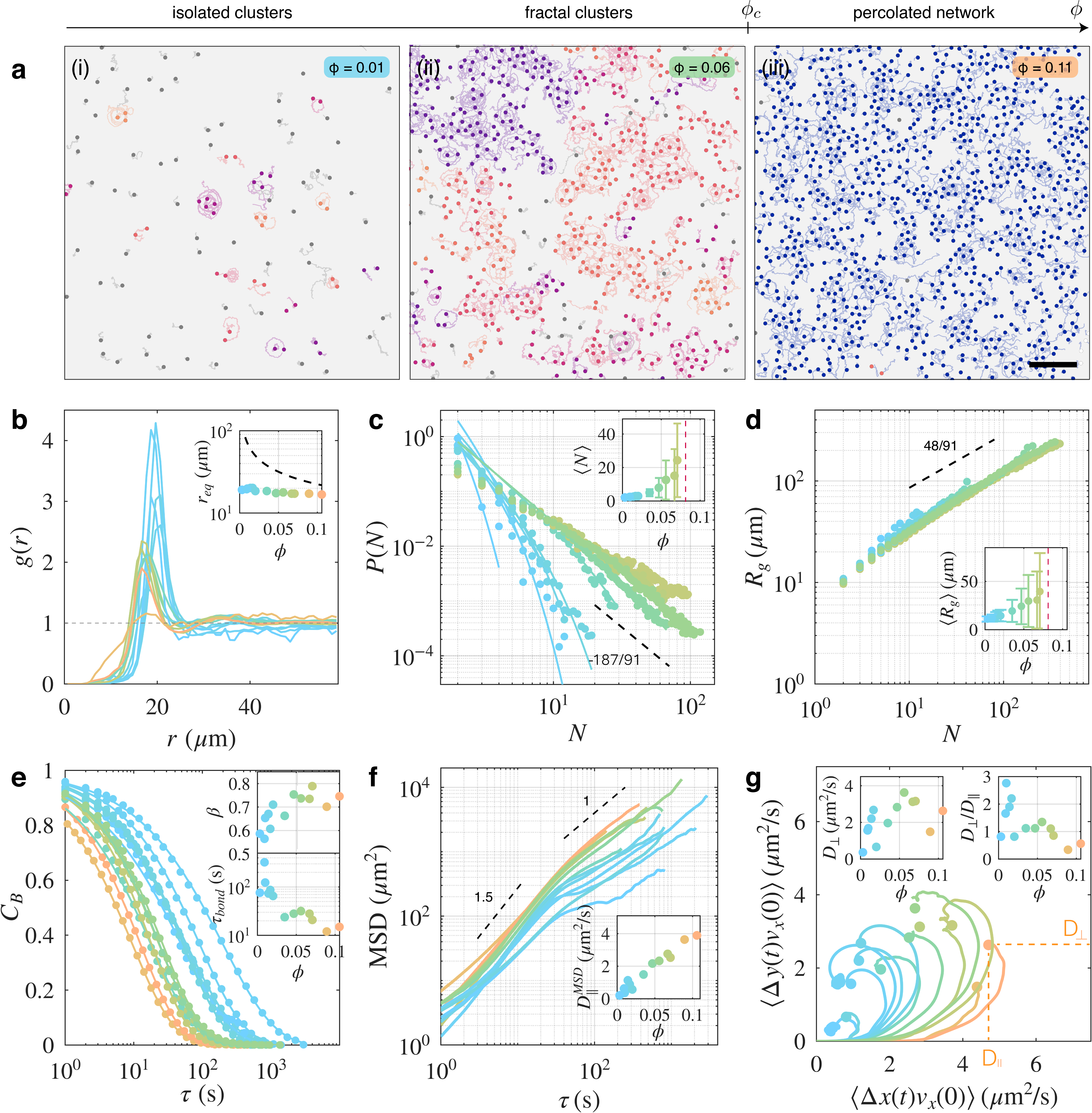} 
	\caption{\textbf{Percolation and odd transport}.
    \textbf{a}: Snapshots of suspensions at increasing packing fractions. Lines show the cells trajectories during maximum \SI{200}{s} and colors denote clusters (gray cells are isolated). Scale bar \SI{100}{\micro\meter}.
    \textbf{b}: Pair correlation function $g(r)$ at different packing fractions. Inset: peak distance $r_{eq}$, with the dashed lines showing the expected interparticle distance for a homogeneous system of similar density.
    \textbf{c}: Cluster size distribution. It evolves from a truncated power law at low $\phi$ (solid lines are fits with $P(N)\sim N^{-\alpha} e^{-N/N^*}$) to a scale-free power law ($P(N)\sim N^{-\alpha}$) close to $\phi_c$. In panels c and d, the dashed line are the critical exponents of 2D percolation theory \cite{staufferIntroductionPercolationTheory1987, nienhuisCriticalBehaviorTwodimensional1984}. Inset: average cluster size, errorbars are standard deviations.
    \textbf{d}: Gyration radius $R_g$ vs. cluster size $N$. It is consistent with the predicted $R_g\sim N^{48/91}$\cite{staufferIntroductionPercolationTheory1987}. Inset: mean $R_g$ vs. $\phi$, errorbars are standard deviations.
    \textbf{e}: Bond survival correlation function $C_B$ (circles), with the fitted stretched exponential $e^{-(t/\tau_B)^\beta}$ (solid line). Insets: fitted exponent $\beta$ and mean bond survival time $\tau_{bond} = \frac{\tau_B}{\beta}\Gamma\left(\frac{1}{\beta}\right)$.
    \textbf{f}: MSD at all packing fractions. Inset: even diffusion coefficient in the linear regime, extracted from the MSD using $\langle \Delta r^2(\tau) \rangle=4D_\parallel^{MSD}\tau$.
    \textbf{g}: Position-velocity correlation functions have a typical logarithmic spiral shape, converging to the even and odd diffusion coefficients $(D_{\parallel},D_{\perp})$ as $t\to\infty$ \cite{hargusOddDiffusivityChiral2021}. To improve statistics, for each curve the $x$- and $y$-coordinates are averages of ($C_{xx}$,$C_{yy}$) and ($C_{yx}$,$-C_{xy}$) respectively, since they are equivalent by symmetry (full plots in SI). Insets: $D_{\perp}$ and thus $D_{\perp}/D_{\parallel}$ are non-monotonic with $\phi$.
    \label{fig:Fig4}
	}
\end{figure}

We next investigate the collective dynamics of suspensions of thousands of spinning microalgae over a wide range of packing fractions $\phi$, where $\phi = A_{\mathrm{cells}}/A_{\mathrm{tot}}$ is the area fraction occupied by the cell bodies. Cells self-organize into highly dynamic rotating clusters: isolated at low $\phi$, they increasingly merge, split and exchange particles as $\phi$ grows, eventually forming an active system-spanning network (Fig.~\ref{fig:Fig4}a-c).
The pair correlation function $g(r)$ shows no positional order beyond nearest neighbors, confirming the fluid-like nature of the system (Fig.~\ref{fig:Fig4}b). Because particles attract each other, the nearest-neighbor distance barely depends on density but is fixed around  $r_{\mathrm{eq}}\approx\SI{20}{\micro\meter}$ (Fig.~\ref{fig:Fig4}b inset). Increasing density therefore results in higher connectivity rather than reduced interparticle spacing. 
Several signatures of this geometrical transition - the cluster size distribution, its fractal dimension, the divergence of the mean finite-cluster size, and the scaling of cluster perimeter with size (SI) - are compatible with the two-dimensional percolation universality class, with a critical density $\phi_c\simeq0.08$ \cite{staufferIntroductionPercolationTheory1987}.

Particle-free regions remain visible even in the dense percolated state (Fig.~\ref{fig:Fig4}a). This empty liquid structure has also been reported in patchy colloids and inertial spinners \cite{bianchiPhaseDiagramPatchy2006}. More broadly, percolation has been observed in other chiral systems with attractive interactions \cite{jorge_order_2026, kushwahaPercolationNonequilibriumAssemblies2024}, suggesting that percolation may be generic when phase separation is frustrated by transverse forces and noise. 

The suspension remains highly dynamic despite the geometrical transition. The bond survival correlation function $C_B(\tau)$, which measures the probability that neighbors present at time $t$ are still neighbors at time $t+\tau$, decays to zero at all $\phi$: the cells' local environment is continually renewed (Fig.~\ref{fig:Fig4}e). It is well described by a stretched exponential with stretching exponent $\beta$ and mean bond survival time $\tau_{bond}$, 
indicating a broad distribution of bond lifetimes \cite{sciortinoOneDimensionalClusterGrowth2005}.
Far from percolation, clusters are small (2-3 particles), isolated, and long-lived ($\tau_{bond}>\SI{100}{s}$). As the density increases, they get bigger, fluidize and frequently exchange particles, causing $\tau_{bond}$ to drop by an order of magnitude ($\tau_{bond}\sim\SI{10}{s}$) and $\beta$ to saturate around $0.8$ - reflecting a local environment that is both more dynamic and more homogeneous on average.

\subsection*{Odd transport on a percolating network}

How does chirality reshape transport? Because the particles break time-reversal and parity symmetries, we expect odd responses (i.e. orthogonal to gradients) from the suspension \cite{hargusFluxHypothesisOdd2025}. We further ask how transport evolves as the cells percolate, extending the classical problem of stirred percolation \cite{laguesElectricalConductivityMicroemulsions1979,bugDiffusionStirredPercolating1987} to systems with transverse interactions.

We first note that, since particles are not self-propelled and remain immobile when isolated, transport is driven entirely by hydrodynamic interactions between the cells. The mean-squared displacement (MSD) shows a crossover from superdiffusive motion at short times, with $\langle \Delta r^2 \rangle \sim t^{1.5}$, to diffusive motion at long times at all packing fractions (Fig.~\ref{fig:Fig4}f). The long-time regime is the classical stirred percolation result: particles hop from cluster to cluster in an effective random walk at the scale of the network \cite{bugDiffusionStirredPercolating1987}. 
However the MSD only reflects ordinary diffusion; to investigate odd effects, we turn to the full diffusivity tensor itself. In two dimensions, it can be written  $D_{ij} = D_\parallel \delta_{ij} + D_\perp \epsilon_{ij}$, with $D_\parallel$ the usual even diffusion coefficient and $D_\perp$ the odd diffusion coefficient driving particle fluxes orthogonal to density gradients. Both can be obtained from Green-Kubo relations, recently generalized to chiral systems: $D_{ij} = \int_0^\infty dt\, \langle v_i(t) v_j(0) \rangle$ \cite{hargusOddDiffusivityChiral2021, hargusFluxHypothesisOdd2025}. In practice, we measure the position-velocity correlations $C_{xx}=\langle \Delta r_x(t) v_x(0)\rangle$ and $C_{yx}=\langle \Delta r_y(t) v_x(0)\rangle$, which converge respectively to $D_\parallel$ and $D_\perp$ at long times \cite{hargusOddDiffusivityChiral2021}. Plotted against each other, $C_{xx}$ and $C_{yx}$ trace characteristic logarithmic spirals centered on $(D_\parallel, D_\perp)$ providing a direct, fit-free measurement of both coefficients at all packing fractions (Fig.~\ref{fig:Fig4}g and SI).

Interestingly, $D_\parallel$ and $D_\perp$ depend differently on $\phi$ and thus on the network topology. While collisions generally hinder diffusion in passive systems, here $D_\parallel$ increases linearly with $\phi$. The underlying mechanism --- neighbors rolling around rather than blocking each other --- is reminiscent of the collision-enhanced self-diffusion recently predicted for odd-diffusive Brownian particles \cite{kalzCollisionsEnhanceSelfDiffusion2022}.
$D_{\perp}$, in contrast, is non-monotonic and reaches a maximum near the percolation threshold. A possible explanation is that, below $\phi_c$, particles have space for transverse motion along cluster boundaries of increasing perimeter, leading to increased edge currents and odd diffusion. Above $\phi_c$, however, the available perimeter length decreases.

The MSD provides further support for this interpretation. Rescaling it by the characteristic cluster size $\langle R_g\rangle$ collapses the plot in space, at all packing fractions where $R_g$ is defined (SI). We therefore interpret the crossover between the two regimes as the transition between particles orbiting within a cluster and hopping to another one. In this picture, the short-time regime probes exploration within a single cluster. In classical stirred percolation, single-cluster exploration is anomalously slow because of the clusters' fractal shape \cite{bugDiffusionStirredPercolating1987}; here, with transverse interactions, it is much faster, consistent with short-time transport being dominated by orbiting motion along cluster boundaries.
Overall, these results show that percolation strongly reshapes both even and odd transport: while even diffusion increases monotonically with connectivity, odd diffusion decreases above the critical percolation threshold, highlighting its distinct sensitivity to network topology.

\subsection*{Discussion}
We introduced a novel experimental system of living rotors that spin clockwise but collectively orbit counterclockwise. This reversal of microscopic and macroscopic chirality differs from previously proposed mechanisms: gear-like coupling through steric interactions \cite{workampSymmetryreversalsChiralActive2018, liaoClusteringPhaseSeparation2018}, boundary effects \cite{beppuEdgeCurrentPairing2021} or dynamical instabilities \cite{hambySwimmingBacteriaPower2018}, which require fine-tuning of density, confinement or activity. Here, the reversal is intrinsic to the cells and directly follows from their internally driven nature, reminiscent of Ref.\cite{markovichChiralActiveMatter2019}. In aligned microswimmers, external driving was shown to simply translate hydrodynamic instabilities in space \cite{vonaHydrodynamicInstabilitiesActive2025, eisenmannPureHydrodynamicInstabilities2025}. Our results show that, in microspinners, the distinction between external and internal driving can qualitatively change collective behavior, from active turbulence \cite{meckeSimultaneousEmergenceActive2023} to chirality inversion and percolation.

Chirality also profoundly reshapes transport. Chiral motion is known to suppress motility-induced phase separation in repulsive active particles \cite{liaoClusteringPhaseSeparation2018, maDynamicalClusteringInterrupts2022}. Here, we observe that phase separation is also frustrated in an odd system with attractive interactions, leading to percolation. We hypothesize that transverse forces continuously break clusters apart, a mechanism that appears generic to chiral active systems \cite{massana-cidArrestedPhaseSeparation2021,  kushwahaPercolationNonequilibriumAssemblies2024, jorge_order_2026}, but  lacks a theoretical description in this regime.
Contrary to the classical picture in which the emergence of a spanning cluster is accompanied by a sudden increase in transport coefficients \cite{laguesElectricalConductivityMicroemulsions1979}, we instead observe that odd diffusion decreases above the percolation threshold. While obstacles have been shown to reshape even diffusion in chiral systems \cite{chanChiralActiveParticles2024, zhangUnusualDiffusionChiral2026, uhligChiralityEnhancesDiffusion2026}, how odd transport responds to geometrical transitions remains largely unexplored.
Our results tie odd transport to the connectivity of the suspension, and raise several questions on the universality of the maximum odd diffusion coefficient at critical packing fraction, or the possible emergence of odd elasticity in dense networks. Further studies on the system introduced here could also shed light on other transport phenomena, such as edge currents or odd viscous effects near free surface flows \cite{franca_odd_2026, plessis_non-reciprocal_2026, metzger_equation_2026}.
Spinning microalgae therefore represent a promising experimental framework for studying how chirality propagates across scales and shapes living systems.



\printbibliography
\pagebreak
\clearpage

\section*{Methods}
\setcounter{figure}{0}
\renewcommand{\thefigure}{S\arabic{figure}}

\subsection*{Cell culture and imaging}

The uniflagellar mutant CC1926 from the Chlamydomonas Resource Center was grown axenically in Tris-acetate-phosphate medium under light/dark cycles (14h/10h). Cells were harvested in the exponential phase and loaded in microfluidic chambers of $\SI{100}{\micro\meter}$ or \SI{400}{\micro\meter} height (Ibidi \textmu-Slide VI 0.1 and \textmu-Slide VI 0.4). When dense suspensions were needed, chambers were kept overnight in the dark. Experiments were imaged at the bottom of the chamber (where both motile and non-motile cells were present) except for dense suspensions, where we imaged cells at the top of the chamber (where almost all cells were motile).
Microscopy movies were acquired with an inverted microscope Nikon TS100 and a Kinetix22 camera. Experiments, except dense suspensions, were imaged in bright-field using objectives ranging from 4× to 30×. Dense suspensions were imaged by spinning-disk confocal microscopy (CrestOptics CICERO).

\subsection*{Cell tracking}

For dilute experiments, the cells orientation was determined by template matching, using a code kindly provided by Louis Woodhams \cite{baylyEfficientSpatiotemporalAnalysis2010}. In short, a cropped image of each cell was manually selected from the first frame of the movie. The initial cell orientation was determined manually by drawing a line along its axis. The cropped image was then used to generate a library of 180 rotated templates. Each subsequent frame was then compared with the rotated template library, and the orientation corresponding to the best match was selected. This simultaneously provided the cell position, allowing the cell trajectory to be reconstructed with great precision (Fig.S3).
For dense experiments, cell positions were identified using StarDist (Fiji). Cell trajectories were then reconstructed using a custom code based on the particle-tracking algorithm of Ref.\cite{crockerMethodsDigitalVideo1996}.

\subsection*{Spinning frequency of individual cells}
To measure the cells individual spinning frequency $\omega$, we first determined their orientation $\theta$ at a high framerate (at least 4 images per rotation). The spinning speed was then obtained by unwrapping $\theta(t)$ and performing a linear fit, with the slope giving the angular velocity.

\subsection*{Distance and orbiting frequencies of cell pairs}
For the orbiting frequency of cell pairs $\Omega$, we first determined the angle between their centers $\psi$ and unwrapped it. As $\psi(t)$ is noisy and non-monotonic, we did not use a linear fit but instead fitted a smoothing spline. Using the \texttt{slmengine} function in MATLAB \cite{derricoSLMShapeLanguage2026}, we directly extracted the smoothed derivative of $\psi$ at \SI{10}{\second} spaced knots. Finally, the distance between cells $r(t)$ was also smoothed with a spline using the same temporal knots, to get rid of fast variations caused by individual cell rotation (Fig.S4)

\subsection*{Flow measurement}

To measure the flow field around single cells, passive tracers were mixed with dilute cell suspensions. We used polystyrene beads of \SI{1.14}{\micro\meter} diameter (PS-FluoRed-Fi329, microParticles GmbH), and imaged isolated cells in bright-field with 90x magnification. We used both low and high framerates, to capture the dynamics of beads far and close to the cell.

To isolate the power and recovery phases of the flagellar beat, we used framerates high enough ($\sim$\SI{200}{\hertz}) to resolve the back and forth rotation of the cells during their beating cycle (Fig.S5). We defined the power phase as the 30\% of the cycle with the highest body angular velocities, and the recovery phase as the 30\% with the lowest angular velocities. This definition was chosen because it provides a consistent threshold independent of natural variations in cell rotation speed. The corresponding instantaneous tracer trajectories were then rotated by $-\theta(t)$, where $\theta(t)$ is the instantaneous orientation of the cell, and averaged to obtain the flow fields in the cell frame - shown in Fig.~\ref{fig:Fig2} of the main text.
The beat-averaged flow field was obtained by similarly rotating all tracer trajectories by $-\theta(t)$ and averaging them over the entire beat cycle.
Finally, the rotation-averaged flow field (i.e., in the laboratory frame) was obtained by directly averaging all tracer trajectories ($>50$ cells), without correcting for the instantaneous cell orientation.

For the instantaneous flow fields (power and recovery strokes), the body shape appeared to modify the flow in the immediate vicinity of the cell. Therefore, we only considered round cells (ellipticity $>0.95$) to ensure a consistent comparison with the far-field model. In contrast, for the rotation-averaged flow field, we included all cells regardless of their shape, as here the goal was to assess whether the model could capture the cells flow and interactions across the full range of observed body shapes.

\subsection*{Comparison between experimental flow and hydrodynamic model}

Our hydrodynamic model of a single-cell consists of two spatially separated forces and a point torque, arranged such that both the net force and torque vanish. The drag and torque are fixed at the origin (the center of the cell body), leaving the thrust magnitude $F$, its location $\ell$, angle $\alpha$ and the distance $h$ between the cell and the wall as free parameters. We manually adjusted these parameters to match the experimental data and the flow map resulting from the addition of these singularities near a wall.

With $a=\SI{5}{\micro\meter}$, we obtained:
\begin{itemize}
    \item $F_{cycle}=\SI{1}{\pico\newton}$, $l_{cycle}=\SI{18}{\micro\meter}$, $\alpha_{cycle}=\SI{126}{\degree}$, $h=\SI{5}{\micro\meter}$.
    \item $F_{power}=\SI{12}{\pico\newton}$, $l_{power}=\SI{15}{\micro\meter}$, $\alpha_{power}=\SI{110}{\degree}$, $h=\SI{5}{\micro\meter}$.
    \item $F_{recovery}=\SI{10}{\pico\newton}$, $l_{recovery}=\SI{8}{\micro\meter}$, $\alpha_{recovery}=\SI{-116}{\degree}$, $h=\SI{5}{\micro\meter}$.
\end{itemize}

The drag and torque values immediately follow from force and torque balance. These estimates are consistent with both \textit{C. reinhardtii} geometry and previous measurements \cite{baylyPropulsiveForcesFlagellum2011, leptosPhototaxisChlamydomonasArises2023}.
Note that the force magnitude is smaller in the rotation-averaged than the instantaneous data. Indeed, the rotation-averaged data also includes the recovery phase of the beat, resulting in an effective force magnitude that is smaller than the instantaneous force during the power stroke.

\subsection*{Cluster analysis}

Cell clusters were identified using the DBSCAN algorithm (MATLAB's \texttt{dbscan}), with a neighborhood radius set to \SI{28}{\micro\meter}, corresponding to the position of the first minimum of the pair correlation function $g(r)$. We verified that cells within this first-coordination shell indeed interacted hydrodynamically, as they were observed to orbit each other. The minimum number of points per cluster was set to 2. To avoid analyzing clusters potentially truncated by the field of view, clusters were excluded from the analysis if more than 40 particles were located within a distance $d_{edge}=\SI{28}{\micro\meter}$ of the image boundaries, or if the cluster contained more of 75\% such edge particles. The clusters perimeter was then defined as the compact boundary perimeter, using MATLAB \texttt{boundary} function with a shrink factor of 1 (Fig.S6).

\subsection*{Logarithmic spirals analysis}

Noting $C_{ij}(t)=\langle \Delta x_i(t)v_j(0)\rangle$, we observed that both $C_{yx}(C_{xx})$ and $C_{xy}(C_{yy})$ trace characteristic spirals, as predicted theoretically \cite{hargusOddDiffusivityChiral2021}. The centers of these spirals correspond to the even and odd diffusion coefficients $D_\parallel$ and $D_\perp$. Having verified that $C_{xx}=C_{yy}$ and $C_{yx}=-C_{xy}$ (Fig.S6d), we exploited these symmetries to average the corresponding components and improve statistics. Specifically, we defined $C_{\parallel}(t)=\frac{C_{xx}(t)+C_{yy}(t)}{2}$ and $C_{\perp}(t)=\frac{C_{yx}(t)-C_{xy}(t)}{2}$, and used these averaged quantities to determine the spirals centers.

Assuming the spirals are logarithmic, their centers can be determined geometrically using the method of tangents --- which we found more reliable than a direct fit. In short, the center of each logarithmic spiral was estimated by finding the point for which the angle between the radial direction (from the center to each point of the trajectory) and the local tangent direction was constant along the trajectory. The center was obtained by minimizing the weighted variance of these angles. Before calculating the tangents, the coordinates were smoothed using a Savitzky–Golay filter. 
The even diffusion coefficient shown in the main text (Fig.~\ref{fig:Fig4}f inset) was obtained by fitting the linear regime of the MSD, using $\langle \Delta r^2(\tau) \rangle=4D_\parallel^{MSD}\tau$. The reported values of $D_\parallel^{MSD}$ are consistent with the estimates of $D_\parallel$ obtained from the $x$-coordinates of the logarithmic spirals centers (shown in SI).

\section*{Data availability}
Experimental data (microscope images and trajectories) of small cell clusters and dense suspensions are available upon request at https://doi.org/10.5281/zenodo.23344033. The full dataset is available upon reasonable request.

\section*{Use of generative AI}
Claude (Anthropic; Claude Opus 5.5 and Claude Fable 5.1) was used to improve grammar and readability, to draft analysis code, and to numerically cross-check the analytical expressions. All AI-assisted content was reviewed, verified and revised by the authors, who take full responsibility for the manuscript.

\section*{Acknowledgements}

We thank Camille Jorge and Antoine Deblais for interesting discussions on odd transport and percolation and for their careful reading of this manuscript, together with Nathan Vani. We also thank Louis Woodhams for kindly providing his template-matching code and tutorial.
We acknowledge support from the ERC Grant No. “2023-StG-101117025, FluMAB.” This publication is part of the Vidi project Living Levers with File No. 21239, financed by the Dutch Research Council (NWO). 

\section*{Author contributions}

I.E.: Conceptualization, Supervision, Methodology, 
Investigation, Data curation, Formal analysis, Writing - original draft, review \& editing.
P.C.: Supervision, Methodology, Formal analysis (derivation of the model), Writing – review \& editing
H.v.S.: Investigation, Data curation, Formal analysis.
T.S.: Investigation, Data curation.
H.F.: Supervision, Methodology, Investigation (numerical simulations), Writing – review \& editing.
M.T.-C.: Supervision, Formal analysis, Writing – review \& editing.
M.J.: Supervision, Funding acquisition, Writing – review \& editing.


\clearpage

\newrefsection

\setcounter{section}{0}
\setcounter{figure}{0}
\setcounter{table}{0}
\setcounter{equation}{0}
\setcounter{page}{1}

\renewcommand{\thefigure}{S\arabic{figure}}

\begin{center}
  {\Large Supplementary Information for\\[4pt]
  ``Chirality inversion and odd transport in a suspension of living rotors"}\\[10pt]
\end{center}

\vspace{0.7cm}

\setcounter{tocdepth}{2}
\tableofcontents

\section{Single cell hydrodynamic model}

\begin{figure}[ht!]
\begin{center}
\includegraphics[width=\linewidth]{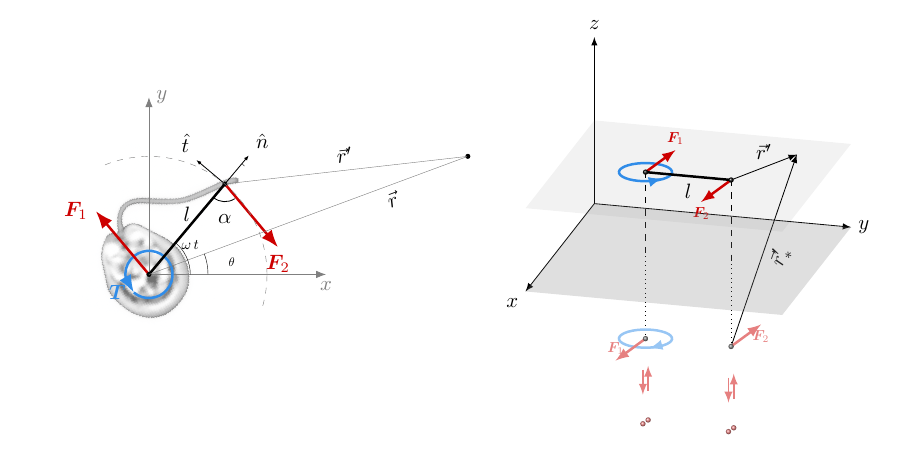}
\end{center}    
    \caption{Geometry of the single cell model, the flow is explained by three singularities. Left --- In-plane view of coordinates and definitions used in the model and corresponding orbital averaging. Note the effective flagellar length $\ell$ and the force angle $\alpha$. Right --- Three dimensional representation of the system of image singularities due to the presence of a wall at a distance $h$ from the swimming plane. The point torque is balanced by another torque of opposite rotation, while the point forces require an additional Stokes doublet and source dipole \cite{blakeFundamentalSingularitiesViscous1974}.}
    \label{fig:method_of_images}
\end{figure}

In order to develop the minimal model for the flow around a single cell, we consider that each individual flow is well approximated when described by the Green's functions of a point torque and two point forces, as schematized in Fig.~\ref{fig:method_of_images} and proposed in \cite{filyCooperativeSelfpropulsionActive2012}. The fast beating of the flagellum allows us to consider a resulting beating force located at an effective flagellar length $l$ with a given orientation; we treat such force as constant. To then balance this pushing force we place the point torque and drag force at the rotational center of cell body. This approach, albeit simple, shows remarkable agreement with the experimental flows retrieved by the tracking velocimetry and importantly captures the observed chirality inversion and the attractive nature of the interaction. However, to fully capture the observed decay, the presence of the wall needs also to be taken into account. 

We considered that the point singularities that describe the single cell flow are coplanar and parallel to a wall at a fixed distance $h$. Thus defining the rotational center $\vec{x}=(0,0,h)$ for the position of the rotlet and with the point forces $\vec{F}=F\vec{e}$ such that $\vec{e}\cdot\hat{\vec{z}}=0$. The image system for the point forces is applied at $\vec{x}_*=(0,0,-h)$ and consists of an opposite point force, a Stokes dipole aligned with the real force, and a source dipole likewise oriented \cite{blakeFundamentalSingularitiesViscous1974}. In the case of the rotlet aligned with the normal to the wall, the image system is simply another rotlet with opposite strength \cite{blakeFundamentalSingularitiesViscous1974}. Further, writing $\vec{r}'$ for the distance vector from the singularity on the $z=h$ plane and $\vec{r}'_*$ the position vector relative to the image system, one has $\vec{r'_*}=\vec{r}'+2h\hat{\vec{z}}$ and $\vec{r'_*}\cdot\vec{e}=\vec{r}'\cdot\vec{e}$, as schematized on Fig.~\ref{fig:method_of_images}. 

Under these conditions, the combined velocity field due to the point force and its image system (i.e., the Blakelet) reads
\begin{equation}
\frac{\vec{u}}{F/8\pi\mu}=\left(\frac{1}{|\vec{r}'|}-\frac{1}{|\vec{r'_*}|}-\frac{2h^2}{|\vec{r'_*}|^3}\right)\vec{e}+\left(\frac{1}{|\vec{r}'|^3}-\frac{1}{|\vec{r'_*}|^3}+\frac{6h^2}{|\vec{r'_*}|^5}\right)(\vec{e}\cdot\vec{r}') \vec{r}'
+\frac{12h^3(\vec{e}\cdot\vec{r}')\hat{\vec{z}}}{|\vec{r'_*}|^5},
\end{equation}
and the rotlet contribution amounts to 
\begin{equation}
        \frac{\vec{u}}{T/8\pi\mu}=\hat{\vec{z}}\times\vec
    {r}\left(\frac{1}{|\vec{r}'|^3}-\frac{1}{|\vec{r'_*}|^3}\right).
\end{equation}
Note that there is an out-of-plane component that is, however, balanced by the counter-rotating movement of the point forces, making its overall effect negligible. In the subsequent discussion we then simply consider the dynamics on the $z=h$ plane.

Due to the rotation of each cell, the position and orientation of the singularities are transported along an approximately circular orbit that we parametrise as
$
\bm R(t)=(R\cos\omega t,\,R\sin\omega t,\,h),
$
and define the moving orthonormal basis
\begin{equation}
\hat{\vec{n}}=(\cos\omega t,\sin\omega t,0),\qquad
\hat{\vec{t}}=(-\sin\omega t,\cos\omega t,0).    
\end{equation}
onto which to project the force director, which we assume describes a solid body rotation along the orbit and has a fixed angle $\alpha$ so $\bm e(t)=\cos\alpha\,\hat{\vec{n}}+\sin\alpha\,\hat{\vec{t}}$.

Let the observation point be $\vec{r}=(r\cos\theta,r\sin\theta,h)$ and thus $\vec{r'}  =\vec{r}-{\bm R(t)}=
(r-R\cos\psi)\, \hat{\boldsymbol{r}}
-
R\sin\psi\,\hat{\boldsymbol{\theta}}$ with $\psi=\omega t-\theta$. Given the separation of time scales between the cell rotation and collective dynamics, we can time-average over one orbit through the phase average
\begin{equation}
\overline{(\cdot)}=\frac{1}{2\pi}\int_0^{2\pi}(\cdot)\,{\rm d}\psi.  
\end{equation}
Performing such an average on the reflected rotlet yields an identically zero radial component  $\bar{\vec{u}}\cdot \hat{\boldsymbol{r}}=0$, as expected,  and for the azimuthal part
\begin{equation}
    \frac{\bar{\vec{u}}\cdot \hat{\boldsymbol{\theta}}}{T/8\pi\mu}
    =\frac{1}{\pi r}\Bigg[ \frac{K(m_R)}{R+r}-\frac{E(m_R)}{R-r}
    -\frac{1}{\sqrt{(r+R)^2+4h^2}}\bigg(K(m_{R}^*)+\frac{r^2-R^2-4h^2}{(r-R)^2+4h^2}E(m_{R}^*)\bigg)
    \Bigg]
\end{equation}
with $m_R=4r R/(r+R)^2$ and $m_{R}^*=4r R/\big[(r+R)^2+4h^2\big]$ and where $K(\cdot)$ and $E(\cdot)$ are the complete elliptic integrals of first and second kind, respectively. If the rotlet happens to be centred, i.e., $R=0$, as we consider in our model, then $m_R=m_{R}^*=0$. Noting that $K(0)=E(0)=\pi/2$, the previous expression simplifies to
\begin{equation}
         \frac{\bar{\vec{u}}\cdot \hat{\boldsymbol{\theta}}}{T/8\pi\mu} =\frac{1}{r^2}\Bigg[ 1
    -\frac{ r^3 }{\left(4 h^2+r ^2\right)^{3/2}}
    \Bigg]=\frac{1}{r^2}\Bigg[ 6\left(\frac{h}{r}\right)^2-30\left(\frac{h}{r}\right)^4+140\left(\frac{h}{r}\right)^6+\cdots\Bigg].
    \label{eq:pedro_model_final_urotlet}
\end{equation}
Following the same strategy for a Blakelet with a co-moving orientation $\alpha$, we obtain
\begin{multline}
\frac{\bar{\vec{u}}\cdot \hat{\boldsymbol{\theta}} }{F/8\pi\mu}   =\frac{2\sin\alpha}{\pi Rr}\Bigg[ \frac{R^2+r^2}{r+R}K(m_R)-(r+R)E(m_R)\\-\frac{1}{\sqrt{(r+R)^2+4h^2}}\bigg({\left(r^2+R^2+4 h^2\right)K(m_R^*)-\big((r+R)^2+4 h^2\big)E(m_R^*)}
\bigg)
\Bigg]   
\label{eq:pedro_model_final_theta}
\end{multline}
for the azimuthal contribution, and 
%
%
\begin{multline}
\frac{\bar{\vec{u}}\cdot\hat{\boldsymbol r}}{F/(8\pi\mu)}
=
\frac{\cos\alpha}{\pi Rr}
\Bigg\{
\frac{R^2+r^2}{r+R}K(m_R)-(r+R)E(m_R)
\\
-\frac{1}{\sqrt{(r+R)^2+4h^2}}
\Bigg[
\left(
r^2+R^2+10h^2
-\frac{h^4m_R^*(2-m_R^*)}{rR(1-m_R^*)}
\right)K(m_R^*)
\\
-\left(
(r+R)^2+4h^2
+\frac{3h^2(2-m_R^*)}{1-m_R^*}
-\frac{2h^4m_R^*({m_R^*}^2-m_R^*+1)}
{rR(1-m_R^*)^2}
\right)E(m_R^*)
\Bigg]
\Bigg\}.
\label{eq:pedro_model_final_radial}
\end{multline}
for the radial part, reusing the same notation for the arguments of the elliptic integrals, $m_R$ and $m_R^*$. Note that the symmetry of the problem imposes that the rotational average of a Blakelet with $R=0$ is identically zero.  

Using Eqs.~\eqref{eq:pedro_model_final_urotlet}-\eqref{eq:pedro_model_final_radial}, we can now build the hydrodynamic model for the cell by adding the contributions from the central rotlet at $R=0$, the Blakelet at $R=0$ and the second Blakelet at $R=l$, as
   %
%
%
\begin{multline}
\bar{\vec{u}}=\bar{\vec{u}}_\text{rotlet}\bigg|_{R=0}+\bar{\vec{u}}_\text{force 1}\bigg|_{R=0}+\bar{\vec{u}}_\text{force 2}\bigg|_{R=l}=\\
u_0\frac{2\sin\alpha}{\pi r}
\Bigg[
\frac{l^2+r^2}{r+l}K(m_l)
-(r+l)E(m_l)
-\frac{\pi l^2}{2r}
\left(
1-\frac{r^3}{(4h^2+r^2)^{3/2}}
\right)
\\
-\frac{1}{\sqrt{(r+l)^2+4h^2}}
\Big(
(r^2+l^2+4h^2)K(m_l^*)
-\big((r+l)^2+4h^2\big)E(m_l^*)
\Big)
\Bigg]\hat{\boldsymbol{\theta}}
\\[8pt]
+
u_0\frac{\cos\alpha}{\pi r}
\Bigg[
\frac{l^2+r^2}{r+l}K(m_l)
-(r+l)E(m_l)
\\
-\frac{1}{\sqrt{(r+l)^2+4h^2}}
\Bigg(
\left[
r^2+l^2+10h^2
-\frac{h^4m_l^*(2-m_l^*)}
{rl(1-m_l^*)}
\right]K(m_l^*)
\\
-\left[
(r+l)^2+4h^2
+\frac{3h^2(2-m_l^*)}{1-m_l^*}
-\frac{2h^4m_l^*({m_l^*}^2-m_l^*+1)}
{rl(1-m_l^*)^2}
\right]E(m_l^*)
\Bigg)
\Bigg]\hat{\boldsymbol r},
\label{eq:pedro_model_final}
\end{multline}

where we introduced the notation $u_0=F/(8\pi\mu l)$ and replaced the rotlet strength by 
$T = -Fl\sin(\alpha)$ to satisfy zero net torque on the cell. The arguments of the elliptic integrals are defined as $m_l=4lr/(r+l)^2$ and $m_l^*=4lr/\big[(r+l)^2+4h^2\big]$.
Eq.~\eqref{eq:pedro_model_final} captures the inversion of chirality in the velocity field through the elliptic integrals branch point at $m_l=1$ that encodes the singularity at $r=l$.

In the far field, where $r$ is much larger than both the effective flagellar length and the distance to the wall, i.e., $r \gg l$ and $r \gg h$ with $h/l = O(1)$, the full expression simplifies to
\begin{equation}
\vec{u}(r,\theta)=
u_0\left[
\frac{12 h^2 l^2}{r ^4}\cos\alpha\,\hat{\boldsymbol{r}}
+
\frac{45h^2l^4}{4 r^6}\sin\alpha\, \hat{\boldsymbol{\theta}}
\right],
\end{equation}
as discussed in the main text.

\section{Numerical model for collective dynamics}
\label{app:chlamys_simulations}

To investigate the collective dynamics of the single-flagellum cells used in the experiments, we will use the minimal particle model introduced in Eq.~\eqref{eq:pedro_model_final}, in which each organism is represented by a point particle carrying the hydrodynamic velocity field generated by its rotating flagellum.

The simulations are formulated in the low-Reynolds-number limit, where the surrounding fluid is described by Stokes flow and the velocity fields generated by individual particles can therefore be superposed. Each particle is advected by the combined hydrodynamic fields of the other particles, while short-range repulsion prevents overlap and stochastic terms account for fluctuations not captured by the deterministic hydrodynamic model. Similar minimal descriptions based on superposed hydrodynamic flow fields have been used to study collective behaviour and interactions between microswimmers \cite{knezevic_collective_2022, lushiPeriodicChaoticOrbits2015, filyCooperativeSelfpropulsionActive2012}.

Rather than attempting a quantitative reproduction of the experimental dynamics, we use this model to identify which ingredients of the particle interactions are most important for destabilising persistent, locked configurations and promoting different collective behaviours. In particular, we investigate the roles of particle-size heterogeneity and stochastic fluctuations associated with close-range interactions, and how these ingredients modify the persistence and rearrangement of rotating particle clusters.

\subsection{Hydrodynamic interaction model}

We denote the position of particle $i$ by $\mathbf{x}_i=(x_i,y_i)$ and its characteristic length scale by $l_i$. The velocity field generated by an individual particle is given by the theoretical Stokeslet-based model introduced in Eq.~\eqref{eq:pedro_model_final}. In nondimensional form, the velocity field generated by particle $i$ can be written as

\begin{equation}
\begin{aligned}
\bar{\mathbf u}_i ={}& \frac{\bar F_i}{\bar L_i}\,\frac{2\sin\alpha_i}{\pi \bar r}\Bigg[
\frac{\bar L_i^2+\bar r^2}{\bar r+\bar L_i}K(m_i) - (\bar r+\bar L_i)E(m_i)
- \frac{\pi \bar L_i^2}{2\bar r}\left(1-\frac{\bar r^3}{\left(4\bar H_i^2+\bar r^2\right)^{3/2}}\right) \\
&\qquad - \frac{\left(\bar r^2+\bar L_i^2+4\bar H_i^2\right)K(m_i^*) - \left[(\bar r+\bar L_i)^2+4\bar H_i^2\right]E(m_i^*)}{\sqrt{(\bar r+\bar L_i)^2+4\bar H_i^2}}
\Bigg]\hat{\boldsymbol\theta} \\
&+ \frac{\bar F_i}{\bar L_i}\,\frac{\cos\alpha_i}{\pi \bar r}\Bigg[
\frac{\bar L_i^2+\bar r^2}{\bar r+\bar L_i}K(m_i) - (\bar r+\bar L_i)E(m_i)\\
& - \frac{1}{\sqrt{(\bar r+\bar L_i)^2+4\bar H_i^2}}\Bigg(
\left(\bar r^2+\bar L_i^2+10\bar H_i^2 - \frac{\bar H_i^4\, m_i^*(2-m_i^*)}{\bar r\bar L_i\,(1-m_i^*)}\right)K(m_i^*) \\
&\qquad - \left((\bar r+\bar L_i)^2+4\bar H_i^2 + \frac{3\bar H_i^2(2-m_i^*)}{1-m_i^*} - \frac{2\bar H_i^4\, m_i^*\left(m_i^{*2}-m_i^*+1\right)}{\bar r\bar L_i\,(1-m_i^*)^2}\right)E(m_i^*)
\Bigg)\Bigg]\hat{\mathbf r},
\end{aligned}
\label{eq:pedro_model_final_ensemble}
\end{equation}
with $m_i = 4\bar L_i\bar r/(\bar r+\bar L_i)^2$ and $m_i^* = 4\bar L_i\bar r/\left[(\bar r+\bar L_i)^2+4\bar H_i^2\right]$.

The nondimensionalization uses the mean particle length $l_\mathrm{avg}=\langle l_i\rangle$ as the characteristic length and $F_\mathrm{avg}=\langle F_i\rangle$ as the characteristic force. Inspecting Eqs.~\eqref{eq:pedro_model_final_theta}--\eqref{eq:pedro_model_final_radial}, these choices naturally lead to $u_{\mathrm{avg}}=F_{\mathrm{avg}}/\left(8\pi\mu l_\mathrm{avg}\right)$  as an option for characteristic velocity.

 The dimensionless parameters associated with particle $i$ are therefore
\begin{equation}
\bar F_i=\frac{F_i}{F_{\mathrm{avg}}},
\qquad
\bar L_i=\frac{l_i}{l_\mathrm{avg}},
\qquad
\bar H_i=\frac{h_i}{l_\mathrm{avg}},
\qquad
\alpha_i.
\label{eq:rescaling_collective_dynamics}
\end{equation}
These parameters allow particle-to-particle variations in force, size, height, and orientation to be incorporated into the model.

A particle $i$ is transported by the superposition of the velocity fields generated by all other particles. Defining the relative position between particles $i$ and $j$ as
\begin{equation}
\mathbf r_{ij}=\mathbf x_i-\mathbf x_j,
\qquad
r_{ij}=\|\mathbf r_{ij}\|,
\end{equation}
the particle positions are updated according to
\begin{equation}
\mathbf{x}_i^{n+1}
=
\mathbf{x}_i^n
+
\Delta t
\sum_{j\neq i}
\left[
\bar{\vec u}_j(\mathbf r_{ij})
+
\bar{\vec{u}}^{\mathrm{rep}}_{ij}
\right]
+
\sqrt{\Delta t}
\left[
\bar{A}_{\mathrm{near}}
\sum_{j\neq i}\boldsymbol{\zeta}_{ij}
+
\bar{A}_{\mathrm{far}}\boldsymbol{\xi}_i
\right],
\label{eq:particle_position_update}
\end{equation}
which corresponds to an Euler--Maruyama time integration \cite{higham_algorithmic_2001}, with the deterministic velocity contributions advanced using a forward Euler step and the stochastic contributions scaled by $\sqrt{\Delta t}$.

The attractive component of the hydrodynamic interaction in Eqs.~\eqref{eq:pedro_model_final_ensemble} and \eqref{eq:particle_position_update} can otherwise lead to an unphysical collapse of the point particles. We therefore include a short-range repulsive velocity based on the Weeks--Chandler--Andersen (WCA) potential \cite{ahmed_phase_2009}. This velocity is given by
\begin{equation}
\bar{\vec{u}}^{\mathrm{rep}}_{ij}
=
\begin{cases}
24\epsilon
\left[
2\left(\dfrac{\bar{d}_{ij}}{\bar{r}_{ij}}\right)^{12}
-
\left(\dfrac{\bar{d}_{ij}}{\bar{r}_{ij}}\right)^6
\right]
\dfrac{\bar{\vec{r}}_{ij}}{\bar{r}_{ij}^2},
&
\bar{r}_{ij}<2^{1/6}\bar{d}_{ij},
\\[8pt]
\mathbf 0,
&
\bar{r}_{ij}\geq2^{1/6}\bar{d}_{ij},
\end{cases}
\label{eq:wca_velocity}
\end{equation}
where the characteristic interaction length is
\begin{equation}
\bar{d}_{ij}=\lambda(\bar{L}_i+\bar{L}_j),
\label{eq:interaction_length}
\end{equation}
with $\lambda$ the dimensionless repulsion length factor. 

We additionally introduce stochastic fluctuations associated with short-range particle interactions, as an experimental cell can be disturbed by collisions with the flagella of its neighbors. The magnitude of these fluctuations is controlled in \eqref{eq:particle_position_update} by the parameter $\bar{A}_{\mathrm{near}}$ and the pairwise stochastic contribution is defined as
\begin{equation}
\boldsymbol{\zeta}_{ij}
=
\begin{cases}
\boldsymbol{\xi}_{ij},
&
r_{ij}<2^{1/6}d_{ij},
\\[6pt]
\mathbf 0,
&
r_{ij}\geq2^{1/6}d_{ij},
\end{cases}
\label{eq:near_noise}
\end{equation}
where $\boldsymbol{\xi}_{ij}$ is a two-dimensional random vector whose components are independently sampled from a standard normal distribution. The second stochastic contribution, $A_{\mathrm{far}}\boldsymbol{\xi}_i$, represents background fluctuations independent of the proximity of other particles, where $\boldsymbol{\xi}_i$ is likewise sampled from a two-dimensional standard normal distribution. Both stochastic contributions are multiplied by $\sqrt{\Delta t}$, corresponding to the scaling of a Wiener process.

\subsection{Polydispersity and initial conditions}

Simulations are performed for populations containing $N$ particles. For each value of $N$, multiple independent realisations are generated by randomising the initial particle positions and particle length scales $\bar{L}_i$. In each realisation, the length values $\bar{L}_i$ are sampled from a log-normal distribution with unity mean and standard deviation $\bar{\sigma}$, which is seen as our controllable polydispersity parameter. We note that the log-normal distribution was observed experimentally for the size of \textit{C. reinhardtii} populations \cite{garz_cell--cell_2012}. Increasing $\bar{\sigma}$ corresponds to increasing heterogeneity in particle size, while the monodisperse case is recovered for $\bar{\sigma}=0$.

The initial particle positions are randomly distributed within a $5\times5$ square in dimensionless coordinates. The particles are therefore generally well separated at the beginning of each simulation, resulting in an initial transient during which the hydrodynamic attraction brings the particles together before a collective rotating state develops.

\subsection{Results: polydispersity or noise are necessary to avoid crystals}

To quantify the spatial organisation of the particles, we calculate the mean pairwise distance (MPD),
\begin{equation}
\bar{D}_{\mathrm{MP}}(t)
=
\frac{2}{N(N-1)}
\sum_{i<j}
\left\|
\bar{\vec{x}}_i(t)-\bar{\vec{x}}_j(t)
\right\|.
\label{eq:MPD}
\end{equation}

The MPD provides a measure of the spatial extent of the particle population. Its initial decrease reflects the aggregation of particles, while a statistically stationary value indicates that the characteristic spatial extent of the population has reached a steady state. The MPD does not distinguish between different geometrical arrangements having similar pairwise separations and is therefore used here as a measure of overall clustering rather than as a complete descriptor of the collective state.

\begin{figure}[t]
    \centering
    \includegraphics[width=\columnwidth]{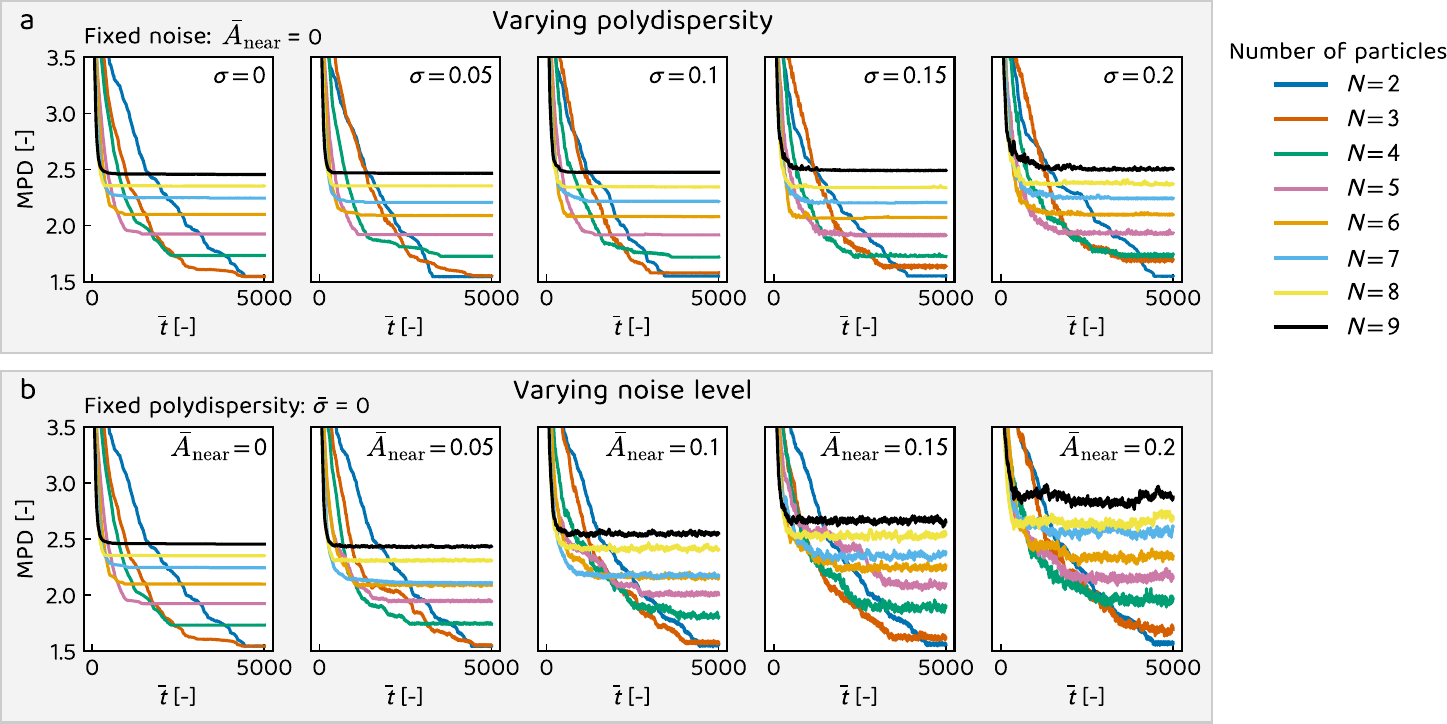}
    \caption{
        \textbf{Collective dynamics numerical simulations.}
        \textbf{(a)} Mean pairwise distance as a function of time for fixed stochastic interaction strength and increasing polydispersity, $\bar{\sigma}=0$, $0.05$, $0.10$, $0.15$, and $0.20$.
        \textbf{(b)} Mean pairwise distance for fixed polydispersity and increasing near-field noise amplitude, $\bar{A}_{\mathrm{near}}=0$, $0.05$, $0.10$, $0.15$, and $0.20$.
        Simulations containing $N=2$--$9$ particles are performed. For each set of parameters $(N, \bar{\sigma}, \bar{A}_\mathrm{near})$, we perform 50 simulations with independently randomised initial positions and particle length scales. Each curve shown in this figure is the average of all 50 individual curves. The time axis is nondimensionalized as $\bar{t} = t \, \cdot \, F_\mathrm{avg}/(8\pi\mu l_\mathrm{avg}^2)$, following the rescaling chosen in Eq.~\eqref{eq:rescaling_collective_dynamics}.
    }
    \label{fig:chlamys_simulations}
\end{figure}

Figure~\ref{fig:chlamys_simulations} summarises the results of the numerical simulations. Each simulation contains between $N=2$ and $9$ particles, and multiple independent realisations are performed for each set of parameters. A full statement of all parameters used in these simulations can be seen in Table~\ref{tab:simulation_parameters}.

\paragraph{Effect of polydispersity.}
In Fig.~\ref{fig:chlamys_simulations}a, the stochastic interaction strength is kept fixed while the polydispersity is varied from $\sigma=0$ to $0.2$. For the monodisperse case, $\sigma=0$, the particles approach one another from their initially separated configuration and subsequently settle into a persistent collective arrangement. Once this configuration is established, the MPD remains approximately constant, indicating that the particles remain locked into a persistent relative configuration. This can also be clearly seen in the supplementary video
, where the monodisperse population becomes locked in a highly stable state.

Increasing the polydispersity does not prevent aggregation, as indicated by the initial decrease of the MPD, but produces persistent fluctuations around the mean MPD at later times. These fluctuations reflect continual relative rearrangements of particles within the rotating cluster, although the cluster itself does not break due to the strong attractive forces.

\paragraph{Effect of stochastic interactions.}
In Fig.~\ref{fig:chlamys_simulations}b, the polydispersity is held fixed while the near-field noise amplitude $\bar{A}_{\mathrm{near}}$ is varied from $0$ to $0.2$. In the absence of stochastic interactions, the particles can settle into a persistent collective configuration. Increasing $\bar{A}_{\mathrm{near}}$ produces increasingly pronounced fluctuations of the MPD about its mean value, qualitatively similar to those obtained by increasing the polydispersity. Not shown here, we also note that increasing this value even further can occasionally temporarily break formed clusters, which also happens in experiments. However, due to the strong attractive force in our model, the cluster eventually reforms unless the background noise $A_\mathrm{far}$ is high enough. The main purpose of the background noise in the model is, therefore, canceling long-range attraction between particles, while the short-range noise is used to break locked clusters.

Together, these simulations show that both particle-size heterogeneity and stochastic fluctuations associated with close-range interactions can disrupt the locked configurations produced by the deterministic hydrodynamic interaction. The model therefore provides a minimal mechanism by which collective rotation can coexist with intermittent particle rearrangements, without requiring the particles to remain indefinitely in a single deterministic configuration.

\begin{table}[ht]
    \centering
    \caption{Parameters used in the multi-particle simulations. The parameters $\alpha_i$ and $\lambda$ are chosen as values measured directly from experiments.}
    \label{tab:simulation_parameters}
    \begin{tabular}{lll}
        \hline
        \textbf{Parameter} & \textbf{Symbol} & \textbf{Value} \\
        \hline
        Number of particles
            & $N$
            & $2$--$9$ \\

        Polydispersity
            & $\sigma$
            & $0$--$0.20$ \\

        Particle lengths
            & $\bar{L}_i$
            & Log-normal distribution, $\langle\bar L_i\rangle=1$ \\

        Dimensionless force
            & $\bar{F}_i$
            & $1$, for all particles \\

        Distance to the wall
            & $\bar{H}_i$
            & $0.25$ \\
        
        Stokeslet angle
            & $\alpha_i$
            & $125^\circ$ \\

        WCA length factor
            & $\lambda$
            & $0.7$ \\

        WCA strength
            & $\epsilon$
            & $0.1$ \\

        Near-field noise amplitude
            & $\bar{A}_{\mathrm{near}}$
            & $0$--$0.20$ \\

        Far-field noise amplitude
            & $\bar{A}_{\mathrm{far}}$
            & $0.01$ \\

        \hline
    \end{tabular}
\end{table}

\newpage
\section{Supplementary figures}

\begin{figure}[h]
	\includegraphics[width=0.9\columnwidth]{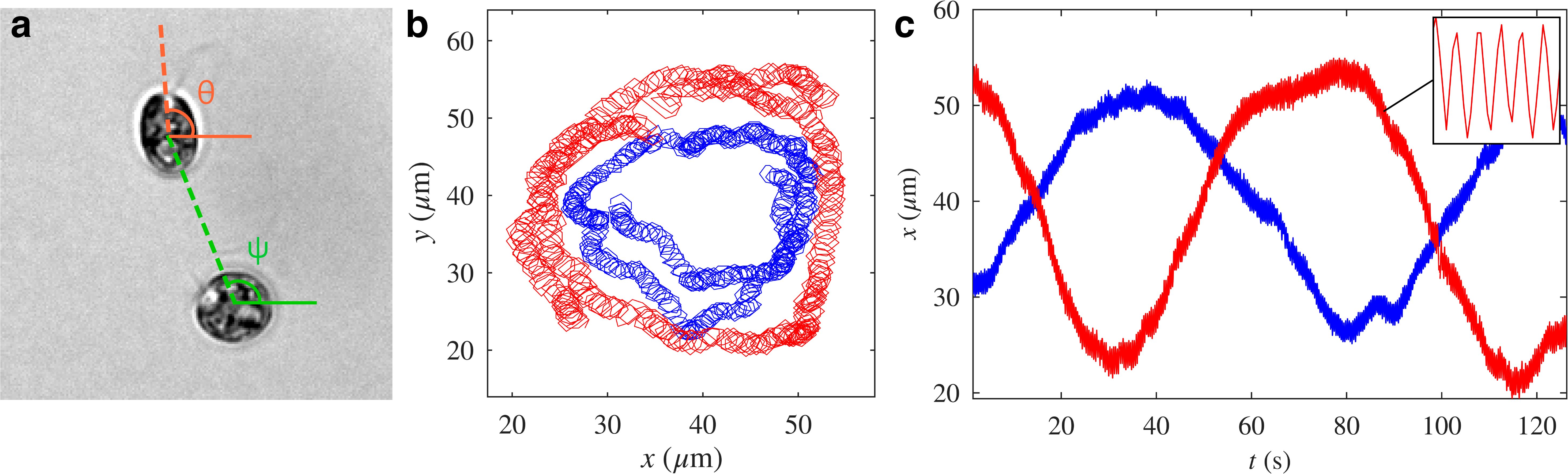} 
	\caption{
        \textbf{Cell tracking.}
        \textbf{a}: Snapshot of two interacting cells, with the cell instantaneous orientation $\theta(t)$, and the angle between both cells $\psi(t)$.
        \textbf{b}: Trajectories of an orbiting cell pair, where red and blue denote each cell. Individual spinning and global orbiting are visible.
        \textbf{c}: Time-evolution of the position $x$ of the same pair. The separation of timescale between spinning and orbiting is clearly visible.
		\label{fig:SI_tracking}
	}
\end{figure}

\begin{figure}[h]
	\includegraphics[width=\columnwidth]{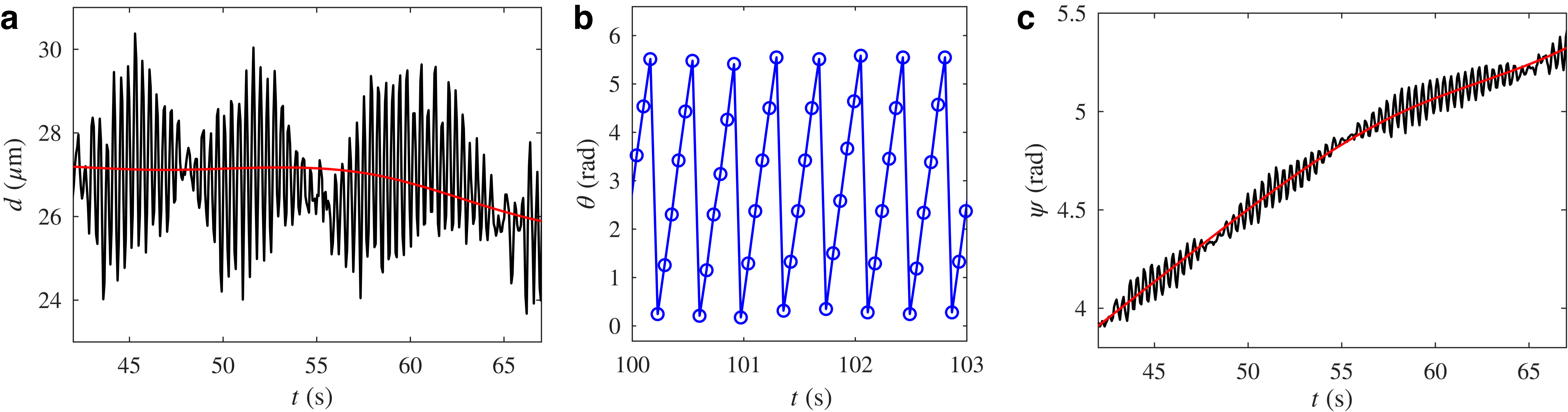} 
	\caption{
    \textbf{Cell pairs analysis.}
        \textbf{a}: Time evolution of the distance $d$ between a pair of interacting cells.
        \textbf{b}: Orientation $\theta$ of a single cell, showing its individual rotation at a constant speed.
        \textbf{c}: Angle $\psi$ between the same orbiting cell pair, showing both the fast spinning and the slower orbiting. In panels a and b, red lines represent the smoothed splines used to compute distance and orbiting speed. Movies were cropped for visualization but are typically $\sim\SI{2}{\min}$ long.
		\label{fig:SI_spin_orb}
	}
\end{figure}

\begin{figure}[h!]
	\includegraphics[width=1\columnwidth]{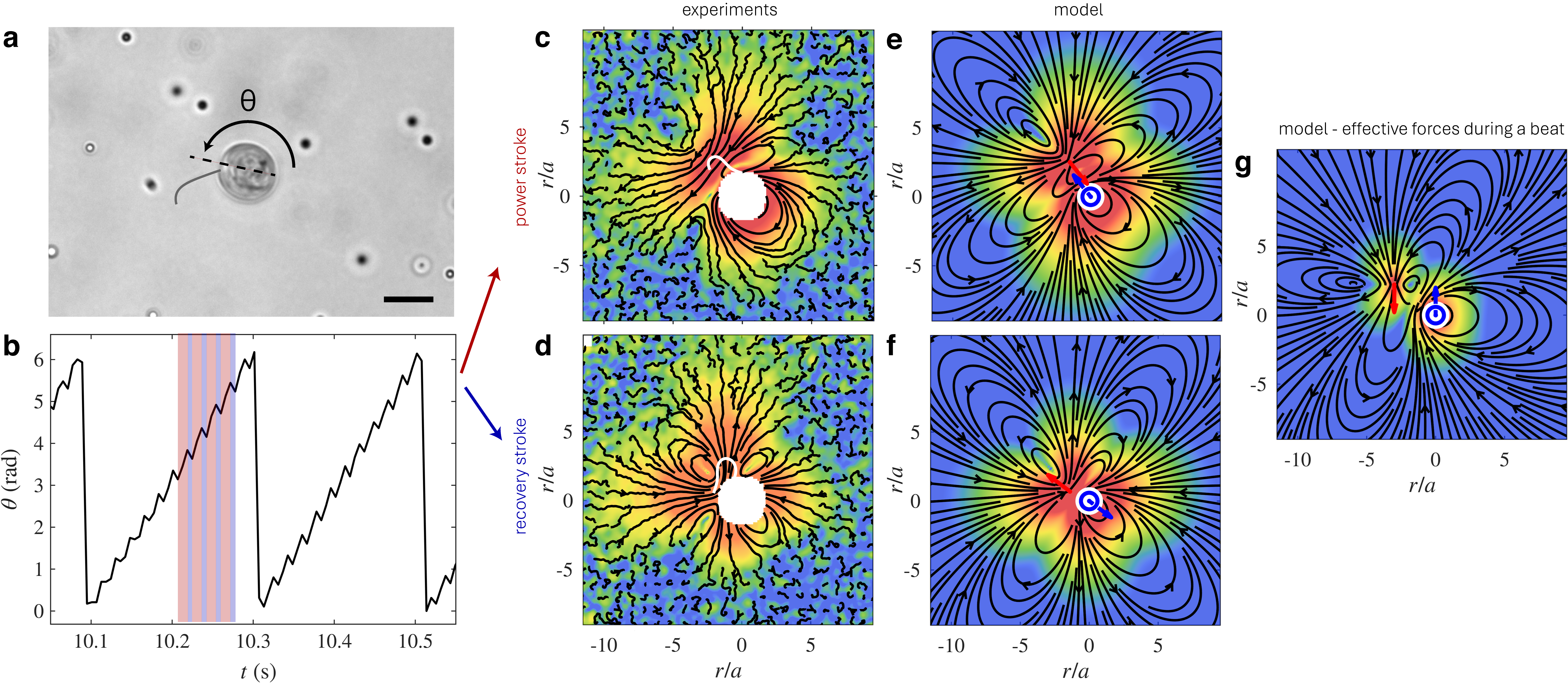} 
	\caption{
        \textbf{Flow measurement.}
        \textbf{a}: Snapshot of a single cell spinning in the presence of tracer beads, with flagellum superimposed for visualization. Scale bar \SI{10}{\micro\meter}.
        \textbf{b}: Example of cell body orientation $\theta(t)$, measured at a high framerate to capture its back-and-forth motion during power (highlighted in red) and recovery (blue) phases of the flagellar beat.
        \textbf{c,d}: Experimental flow field during the power (30\% highest body angular velocities) and recovery (30\% lowest body angular velocities) phases of the beat respectively, reconstructed in the cell frame considering only round cells of ellipticity$>0.95$.
        \textbf{e}: Model of a 2D active rotor near a wall, fitted manually to the power phase: $F_{power}=\SI{12}{\pico\newton}$, $l_{power}=\SI{15}{\micro\meter}$, $\alpha_{power}=\SI{110}{\degree}$, $h=\SI{5}{\micro\meter}$. Arrows denote forces (drag and thrust) and the circle the torque.
        \textbf{f}: Model of a 2D active rotor near a wall, fitted manually to the recovery phase: $F_{recovery}=\SI{10}{\pico\newton}$, $l_{recovery}=\SI{8}{\micro\meter}$, $\alpha_{recovery}=\SI{-116}{\degree}$, $h=\SI{5}{\micro\meter}$.
        \textbf{g}: Model of a 2D active rotor near a wall, fitted to the effective beat-averaged flow field (manually fitted to the experimental rotation-averaged flow field of Fig.2a): $F_{cycle}=\SI{1}{\pico\newton}$, $l_{cycle}=\SI{18}{\micro\meter}$, $\alpha_{cycle}=\SI{126}{\degree}$, $h=\SI{5}{\micro\meter}$.
	}
    \label{fig:SI_flow}
\end{figure}

\begin{figure}[h!]
    \centering
	\includegraphics[width=1\columnwidth]{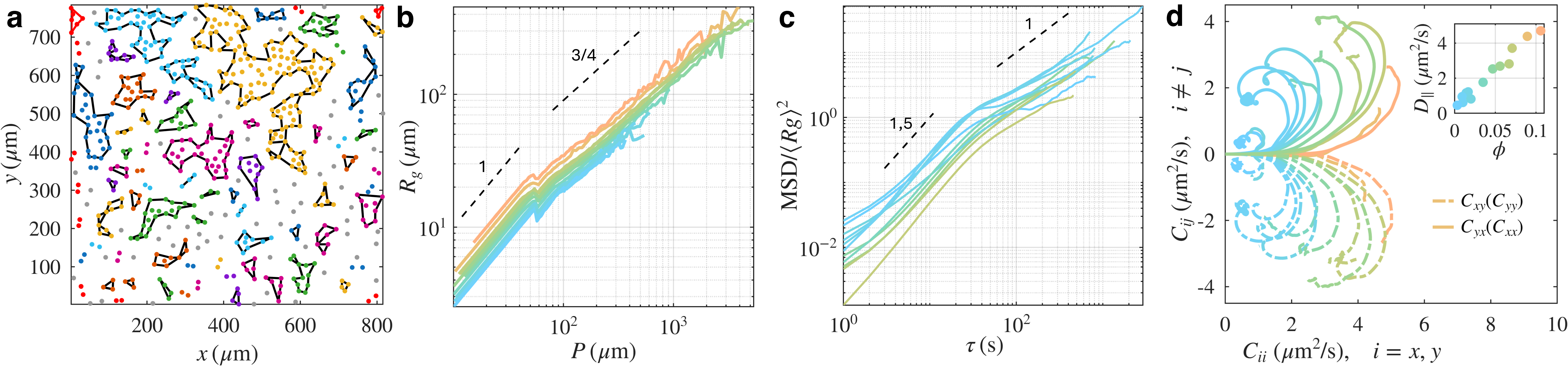}
	\caption{
        \textbf{Clusters analysis and position-velocity correlation functions.}
        \textbf{a}: Example snapshot of particle clusters, with solid line denoting each cluster's perimeter, as used in panel b. Colors denote clusters; red particles are discarded from the analysis due to their proximity with the field-of-view edges.
        \textbf{b}: Radius of gyration of the clusters $R_g$ as a function of the clusters perimeter $P$. The slope 1 indicate linear or circular clusters, which perimeter increases linearly with radius, and 3/4 is the scaling expected from fractal clusters in 2D percolation theory. Colors denote cells packing fractions; plots are shifted for visualization. 
        \textbf{c}: MSD rescaled by the averaged cluster radius of gyration $\langle R_g\rangle$, as a function of the lag time $\tau$.
        \textbf{d}: Position-correlation functions, with $C_{ij}(t)=\langle \Delta x_i(t)v_j(0)\rangle$. Inset: even diffusion coefficient corresponding to the $x$-coordinate of the logarithmic spirals center.
	    }
    \label{fig:SI_cluster}
\end{figure}


\newpage
\section{Caption of supplementary movies}

\textbf{Supplementary Movie 1}: Single cell spinning in the presence of tracer beads, shown first 20x slower, then 3x slower than real time, to illustrate both the flagellar beat cycle and the spiral flow it generates.

\textbf{Supplementary Movie 2}: Time-evolution of the flow field generated by an isolated spinning cell over one flagellar beat cycle, reconstructed in the cell reference frame. The beat phase is inferred from the instantaneous rotation rate of the cell body, which rises during the recovery-power stroke phases, and falls during the power-recovery phases. 33 movies are rescaled by the cell body radius $a$ before averaging.
All cells were included regardless of body shape, to maximize statistics specifically around the power-recovery stroke transition which is very fast.

\textbf{Supplementary Movie 3}: Pair of interacting cells. Movie shown first $2.5\times$ slower, then $4\times$ faster than real time to illustrate both the cells individual CW spinning and pairwise CCW orbiting.

\textbf{Supplementary Movie 4}: Movies of dense suspensions at different packing fractions.    

\textbf{Supplementary Movie 5}: Simulations of collective motion of spinning particles in Stokes flow. In the top row, no noise is applied to the model, and each panel corresponds to a different amount of polydispersity. In the bottom row, the particles are monodisperse, and each panel correspond to different levels of noise. We see that both polydispersity and noise can break the highly stable states that naturally arise with the hydrodynamic model.

\printbibliography[title={Supplementary References}]

\end{document}